# Phase-transitions in spin-crossover thin films probed by graphene transport measurements

J.Dugay,*,† M.Aarts,† M.Giménez-Marqués,‡,¶ T.Kozlova,† H.W.Zandbergen,†
E.Coronado,*,‡ and H.S.J van der Zant†

*Kavli Institute of Nanoscience, Delft University of Technology, Lorentzweg 1, 2628 CJ Delft, The Netherlands., Instituto de Ciencia Molecular (ICMol), Universidad de Valencia, c/ Catedrático José Beltrán, 2, 46980 Paterna, Spain., and Institut Lavoisier, UMR CNRS 8180, Université de Versailles Saint-Quentin-en-Yvelines, 45 Av. des Etats-Unis, 78035 Versailles cedex, France.*

E-mail: julien.dugay@gmail.com; Eugenio.Coronado@uv.es

**Abstract**

Future multi-functional hybrid devices might combine switchable molecules and 2D material-based devices. Spin-crossover compounds are of particular interest in this context since they exhibit bistability and memory effects at room temperature while responding to numerous external stimuli. Atomically-thin 2D materials such as graphene attract a lot of attention for their fascinating electrical, optical, and mechanical properties, but also for their reliability for room-temperature operations. Here, we demonstrate that thermally-induced spin-state switching of spin-crossover nanoparticle thin films can be monitored through the electrical transport properties of graphene lying underneath the films. Model calculations indicate that the charge carrier scattering mechanism in graphene is sensitive to the spin-state dependence of the relative dielectric constants of the spin-crossover nanoparticles. This graphene sensor approach can be applied to a wide class of (molecular) systems with tunable electronic polarizabilities.

Maintaining the unique magnetic, optical and mechanical properties of multifunctional-molecular materials at the nanoscale in synergy with solid-state device structures is an appealing alternative to silicon in post-CMOS devices.[1] Multi-functionalities arising from spin-crossover (SCO) materials have enjoyed attractiveness in recent years[2–4] because these materials favour either a high spin (HS) or low spin (LS) state, which can be switched through various external stimuli from cryogenic up to temperatures well above ambient conditions. Upon spin-state switching, significant changes in the metal-ligand bond length and geometry occur, as well as in their molecular volume.[5–8] A rich variety of switching behaviors arises from the engineering of elastic interactions between SCO molecules, exhibiting gradual, first-order step transitions or even hysteresis effects. Furthermore, this modification of the vibrational modes comes along with a change in color, magnetic susceptibility,[9] mechanical properties,[10] as well as in electrical conductance and dielectric constant.[11] As such, SCO materials have been proposed for integration into displays,[12] memory devices,[9] molecular spintronic devices,[13] pressure and temperature sensors,[14] gas sensors,[15] nanothermometers,[16] optoelectronic devices[17] and actuators.[18]

SCO materials have been considered as good candidates for active elements in electronic devices after the discovery of the thermal bistability in the electronic polarizability of several SCO compounds.[19] For example, a large memory effect in the electrical conductivity has been reported while investigating the [Fe(Htrz)$_2$(trz)](BF$_4$) (Htrz = 1H-1,2,4-triazole) SCO compound in form of bulk powders[20,21] and micro-materials.[22–26] Subsequently, nano-electronic devices with spin-state switching functionality preserving the desirable bulk properties have been reported.[13,22,27,28] However, even though a robust memory effect in the conductance has been recently achieved thanks to the presence of a silica shell,[28] practical operations remain hindered due to the insulating character of these SCO nanomaterials.

Here, we combine a single-layer graphene device with bistable thin-films made of SCO Fe(II) nanoparticles. We measure the graphene conductivity in a four-probe field-effect configuration as a function of temperature and gate voltage. We found that charge carriers in the graphene sheet are affected by the spin-state, allowing to probe the spin-transition and associated memory effect. The change in the scattering originates from the spin-state dependent switching of the electronic polarizability of the SCO nanoparticles covering the graphene, as suggested from model calculations. This simple route of sensing the spin-state, demonstrated for two SCO nanoparticle systems, is thought to work for nanoscale detection of numerous other SCO complexes as the relative dielectric permittivity always changes upon spin-transition.

## Results

A schematic representation of the graphene based field-effect device is shown in Fig. 1a. It consists of large-area chemical vapour-deposited (CVD) graphene on a Si/SiO$_2$ substrate, on top of which a thin film of [Fe(Htrz$_2$)(trz)](BF$_4$)[29,30] spin-crossover nanoparticles has been deposited by a $\mu$-contact printing technique (see methods). This SCO nanoparticle system, hereafter referred to as system (1), has an average length of 25 nm along the rod direction and an average diameter of 9 nm (see Ref 27 and Supplementary Fig. 6a and 7). In order to demonstrate that this approach is general, a spin-crossover thin film for a different chemical alloy [Fe(Htrz)$_{2.95}$(NH$_2$trz)$_{0.05}$](ClO$_4$)$_2$[29,30] nanoparticle system was prepared under the same conditions. As revealed by TEM characterizations in Supplementary Fig. 6b and 7, this SCO nanoparticle system possesses an average diameter of 10 nm with a round shape (hereafter referred to as system (2)). This second system exhibits lower critical temperatures and a narrower hysteresis cycle, as it is expected for the amino-triazole ligand substitution effect. The SCO systems (1) and (2) are both made of a spin-crossover Fe(II) core coated with a surfactant shell. A representative optical microscopy image taken after nanoparticle deposition of system (1) is shown in Fig. 1b. The obtained results in terms of film thickness and homogeneity are in good agreement with previous results,[13,27] indicating that a large-area thin-film has been obtained (see optical and AFM characterizations of the SCO system (2) in Supplementary Fig. 8). The two investigated SCO systems present thermally induced phase-transitions accompanied by a memory effect above and near room temperature respectively, as evidenced by magnetization data obtained from powder samples (see Supplementary Fig. 9).

The resistance of the graphene sheet as a function of temperature and gate voltage, R(V$_G$,T), was measured in a four-probe contact configuration, before and after decoration of the device with a monolayer of spin-crossover nanoparticles. Four-probe measurements were carried out in a van der Pauw configuration inside a probe-station, by placing the probes through the nanoparticle film on the graphene covered sample. Current is injected through two probes on one side, and the voltage is measured across the other two probes to obtain current-voltage characteristics (IV's) for different temperatures and gate voltages, as illustrated in Fig. 1.

Before printing the nanoparticle layer, the electrical properties of as-purchased graphene chips were characterized as a reference using a four probe contact configuration (see methods). In a cleaning step, the sample was heated while flushing nitrogen gas. The annealing temperature was limited to the value at which the highest switching temperature is expected ($\sim$ 380 K for the nanoparticle system (1)). Figure 2a presents the back-gate dependence of the graphene sheet resistance during several cleaning steps. Dots and lines correspond respectively to the data and model fits. The latter have been performed following a diffusive transport field-effect model[31,32] to extract device characteristics such as charge carrier mobilities, maximum resistance and voltage shift at the Dirac point (see methods section for more details). The obtained values are summarized in the Supplementary Table 1. After flushing N$_2$ gas into the chamber at room temperature, the standard p-doped ambipolar field-effect of graphene is observed. We subsequently annealed the sample for several hours under N$_2$ gas up to $\sim$380 K. The first annealing step led to a shift of the Dirac point to lower voltages, and a decrease of the Dirac point resistance, as well as an increase in the charge carrier mobilities (see the evolution of the whole process in Supplementary fig. 10). Figure 2a also illustrates that the graphene sheet remains p-doped after a second annealing step, as the Dirac point stabilizes at a positive gate voltage of 28 V.

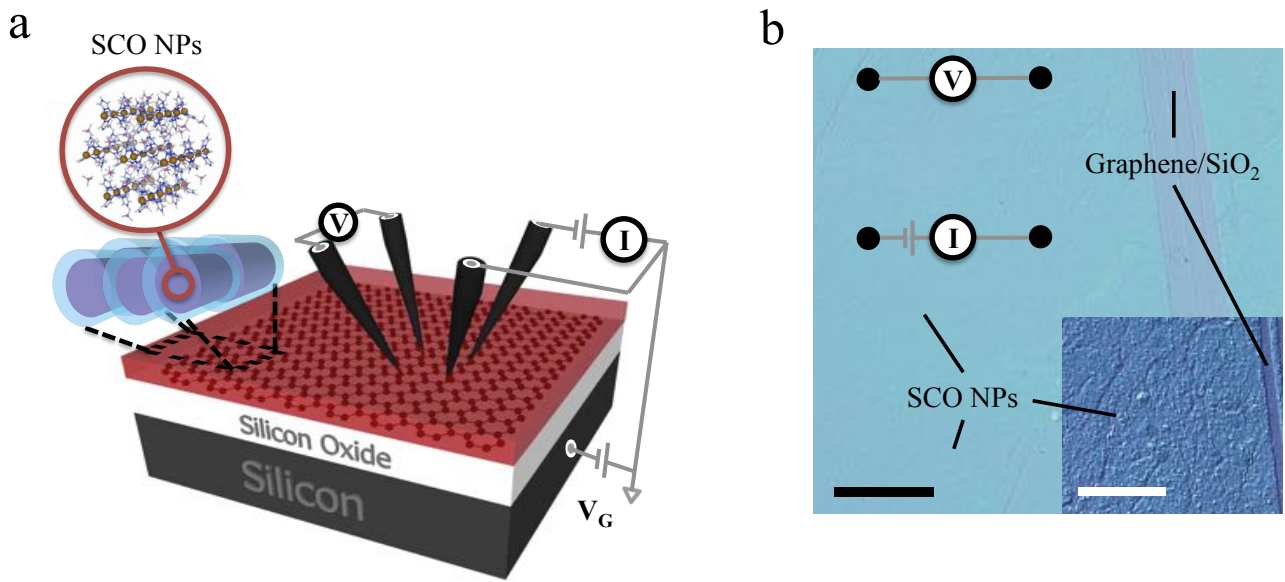

Figure 1: (**a**) Schematic of the device with (CVD) graphene on top of a silicon-silicon oxide substrate and after deposition of a bistable SCO nanoparticle thin film prepared by $\mu$-contact printing. The transport properties of graphene have been determined from the four-probe sample resistance in a van der Pauw configuration. (**b**) Optical image of the device after deposition of a spin-crossover nanoparticle monolayer, scale bar: 200 $\mu$m (Inset: zoom in on the deposit evidencing its surface roughness with a scale bar of 50 $\mu$m). The stripe on the right side is a scratch in the film, and serves for the purpose of graphene/SiO$_2$ reference in this image. The four black circles sketch the location of the four probes and thus the area probed with dimensions L = W = 350 $\pm$ 25 $\mu$m.

Figure 2b displays the graphene sheet resistance of the reference sample as a function of the temperature at zero gate voltage. These data were obtained, after pre-conditioning of the device. The positive temperature dependence of the resistance in heating and cooling modes (respectively solid and open blue dots), demonstrate the metallic behavior of the reference sample. This metallic behavior holds for all backgate voltages as shown in the Supplementary Fig. 10 and Supplementary Table 1. A representative linear voltage (V) vs. current (I) curve is shown in the inset of this figure suggesting an ohmic contact for the electrodes on graphene (all current-voltage characteristics are displayed in the Supplementary fig. 11). More importantly, figure 2b shows that no hysteresis in the resistance was observed before deposition of the SCO nanoparticles.

Figure 3 presents electrical properties of a graphene device after its decoration with the system (1) ([Fe(Htrz)$_2$(trz)](BF$_4$)) SCO nanoparticles. The data have been recorded after the pre-conditioning protocol was applied to this device, consisting of seven thermal cycles carried out from room temperature up to 380 K (see the pre-conditioning data of the SCO system (2) in Supplementary Fig. 12). It should be noted that the Ohmic contact is maintained after deposition of the nanoparticles (see Supplementary Fig. 13 for SCO nanoparticles of the system (2)).

The top and bottom panels of Fig. 3a correspond to the graphene resistance as a function of the backgate voltage in the heating (top) and cooling (bottom) modes. Data and associated fits, performed for the whole temperature range, are depicted by the full and dashed lines, respectively. The back-gate was ramped between 0 and 111 V, set to prevent gate oxide breakdown. The fits are plotted for gate voltages up to 220 V. It is interesting to note that the Dirac point voltage and resistance increased significantly after deposition. Figure 3b shows the temperature dependence of the resistance for specific backgate voltages in the range of 0-111 V. One can observe a pronounced decrease (increase) of the resistance above (below) the switching temperatures of 374 (327 K) in the heating (cooling) mode in the top (bottom) figures. No back-gate dependence of these switching temperatures was noticeable as illustrated in Supplementary Fig. 14. In addition, it should be noted that the underlying slope of the R(T)-curve changes as a function of the gate voltage.

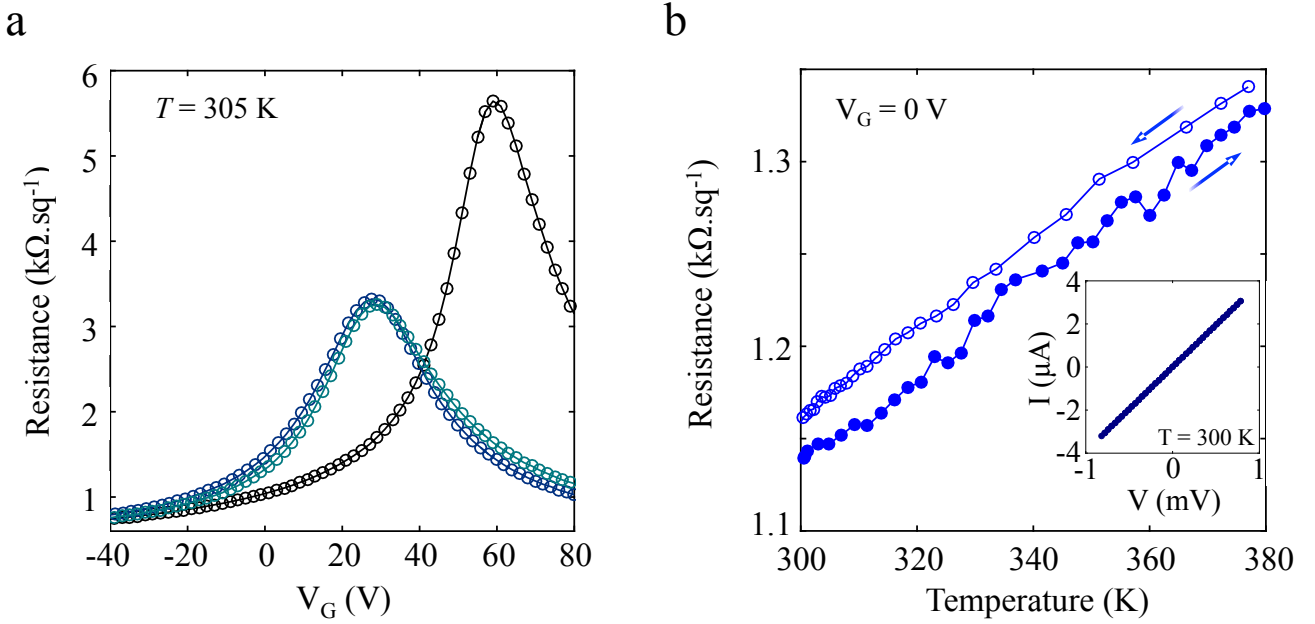

Figure 2: (**a**) Gate dependence of the graphene resistance per square recorded at room temperature before SCO nanoparticle deposition. Dots and lines correspond respectively to data and fits, from which charge carrier mobilities, Dirac point resistance and voltage values were extracted, summarized in Table 1. The standard ambipolar field-effect of graphene is observed after $N_2$ flushing at room temperature (black dots). After annealing for several hours under $N_2$ gas up to ∼380 K, the Dirac voltage shifts towards lower gate voltages; the resistance at the Dirac point decreases and the mobility improves. These trends stabilized after a subsequent annealing step (blue dots). (**b**) Temperature dependence of the resistance in heating and cooling modes (full and empty blue dots, respectively) at zero gate voltage recorded after the graphene was cleaned. The inset shows the representative behavior of the voltage (V) as a function of the injected current (I) indicating an ohmic contact for the electrodes on graphene (see all current-voltage characteristics in Supplementary Fig. 11).

At low voltages, the decorated graphene sheet displays metallic behaviour (with dR/dT ≥ 0) for the whole temperature range. In contrast, this changes for gate voltages near the Dirac point where the slope becomes flat or even displays an insulating-like character (with dR/dT < 0).

The temperature dependence of the Dirac point resistance is presented in Fig. 4a. A well pronounced clockwise hysteresis loop is visible in the data at the same critical temperatures (374 K and 327 K) as the ones observed in the resistance measured at different backgate voltages (see Fig. 3). A consistent clockwise hysteresis loop for the SCO nanoparticle system (2) was also observed (see the Supplementary Fig. 12b). It is important to note that at the same time, only a slight monotonous increase of the Dirac point gate voltage as a function of temperature is observed (see the Supplementary Fig. 16). In other words, there is no dependence of the Dirac point gate voltage on the spin state of the SCO thin film.

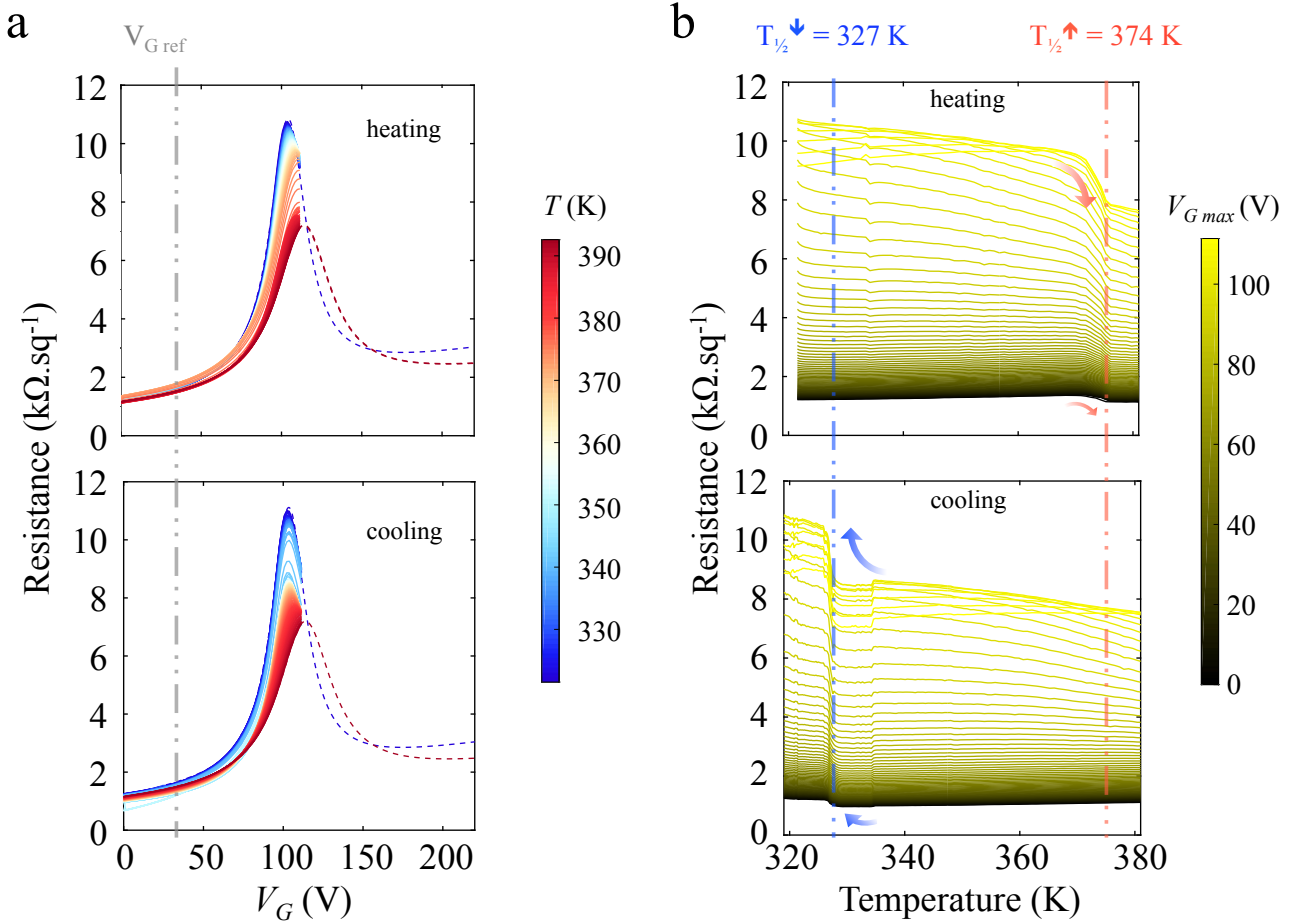

Figure 3: (a) Graphene resistance versus backgate voltage ($V_G$) after deposition of nanoparticle system (1) as a function of heating and cooling modes (top and bottom graphs, respectively). Drawn and dashed lines represent data and fits (the latter shown for clarity only for the extreme temperature values) for a back-gate ramped between 0-111 V and 0-220 V, respectively. $V_{G\,ref}$ indicates the Dirac point of the undecorated reference positioned at $V_G = 28$ V. (b) Resistance versus temperature curves extracted for different backgate voltages in the range of 0-111 V, *i.e.,* from far to close to the Dirac point. A pronounced decrease (increase) of the resistance above (below) 374 (327 K) occurs in the heating (cooling) mode (top and bottom graphs, respectively) (see also main text).

Figure 4b shows the temperature dependence of the field-effect hole mobility for the heating and cooling modes extracted from the steepest regime in the resistance vs. backgate voltage data presented above in Fig. 3a. The hole mobility of graphene after decoration is almost halved compared to before deposition of the nanoparticles (see Supplementary Fig. 10 and Supplementary Table 1 for numeric values). The electron mobility could not be determined, as the Dirac point was close to the maximum experimental gate voltage, bringing the part of the gate dependence with negative slope out of the experimental range. The first important feature that can be observed is the well-pronounced anticlockwise hysteresis loop in the hole mobility. Moreover, the critical temperatures of these loops are consistent with the ones observed in the temperature-dependent resistance curves (see Fig. 3b and 4a). The second important feature is the continuous decrease of the mobilities with temperature. This behavior is an intrinsic and known electronic feature for bare graphene,[33] consistent with what we observed for graphene before deposition (see Supplementary Table 1).

The most natural explanation for the hysteresis loop observed in the resistance and carrier mobilities of the graphene layer is that it is connected to the spin transition of the decorating SCO thin film. This conclusion is strengthened by the absence of the same feature in the reference sample. The main question is then how the spin state of the SCO nanoparticles couples to the transport properties of the graphene layer. Generally speaking, there are two main mechanisms that can cause resistance changes after graphene decoration: *i)* charge-transfer to or from the nanoparticles, or *ii)* a change in the scattering of charge carriers in the graphene due to the presence of the particles on top. Spin-state dependent charge transfer is unlikely because of the absence of switches in the Dirac voltage as a function of temperature. The slight shift of the Dirac voltage with temperature resembles the one observed for bare single-layer graphene studied by Zhu *et al.*[33] We thus conclude that the presence of the SCO thin film predominantly affects the charge carrier scattering in the graphene sheet.

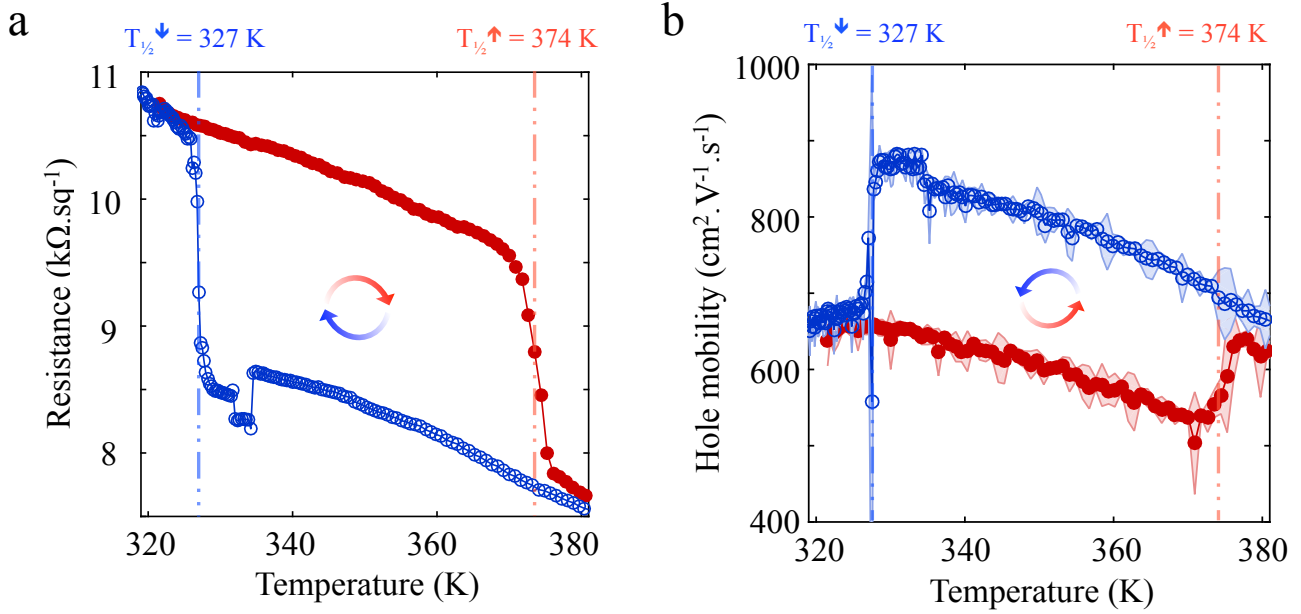

Figure 4: Temperature dependence of graphene electrical properties after deposition of system nanoparticle (1). (**a**) Resistance per square at the Dirac point as a function of temperature for heating and cooling modes. A wide clockwise hysteresis loop is observed between 327-374 K. (**b**) Hole mobility *vs.* temperature for the heating-, and cooling mode. Shaded error bars represent experimental errors in determining the field-effect mobility (see the Methods section for fitting procedure).

In the temperature range used in this work (300-380 K), the main scattering contribution in pristine graphene has been attributed to (thermally activated) remote interfacial phonons (RIP): optical phonons in the $SiO_2$ substrate scatter the carriers in graphene by means of a fluctuating electric field near the interface.[34,35] In this vein we suggest RIP scattering at the nanoparticle/graphene interface as the dominant factor influencing the change in carrier mobilities and resistance in our device. There are two possible origins for the coupling between the SCO in the thin film and the graphene conductivity. Firstly, the vibrational modes in the SCO complexes are different for both spin states, possibly resulting in a change in the contribution to the scattering due to phonons originating from the SCO thin film. Secondly, it is known that the dielectric constant of the $[Fe(Htrz)_2(trz)](BF_4)$ SCO polymeric material decreases substantially from the low-spin to the high-spin states.[36] This is expected to affect the coupling of the phonon modes due to a changing polarizability of the interface.[37] To provide a better understanding of the different factors contributing to the RIP scattering rate of the charge carriers, we modified the model from Chen *et al.*[35] describing the gate- and temperature dependence of graphene on $SiO_2/Si$:

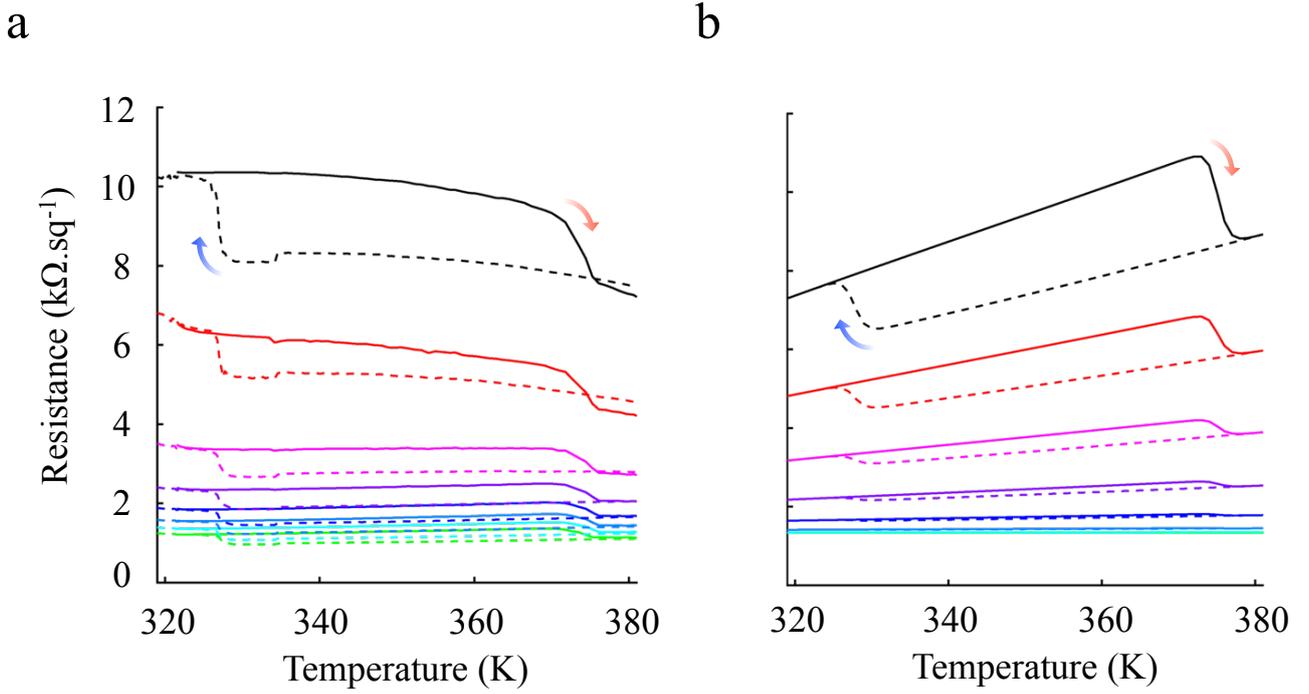

Figure 5: Comparison between model calculations and experimental results of the memory effect in the resistance versus temperature. (**a**) Data of graphene resistance versus temperature selected from Fig. 3b for 8 different backgate voltages ranging from 0 to 110 V for comparison to (**b**) calculated spin-state dependence of the resistance taking into account the main phonon modes and dielectric properties of the spin-crossover nanoparticles, and their respective coupling to graphene charge carriers. The model captures the main features of the data, including the presence and the clockwise direction of the hysteresis loops. The solid and dashed lines indicate heating and cooling mode, respectively. The parameters used in the model are $\alpha$ = 3.5, $B_1$ = 4.25 × $10^{-5}$(h/$e^2$), $B_3$ = 1.5 × $10^{-6}$(h/$e^2$) and d = 1.5 nm. The low-temperature resistivity $\rho_0$ = 1.3 k$\Omega$ has been estimated by matching the experimental resistance measured at room temperature and high carrier density.

$$\rho(V_G,T) = \rho_0(V_G) + \rho_A(T) + \rho_B(V_G,T), \text{with } \rho_B(V_G,T) = \rho_{B\ SiO_2}(V_G,T) + \rho_{B\ NP}(V_G,T), \quad (1)$$

and

$$\rho_{B\ SiO_2}(V_G,T) = B_1 V_G^\alpha \left( \frac{1}{e^{\frac{59\ meV}{k_B T}} - 1} + \frac{6.5}{e^{\frac{155\ meV}{k_B T}} - 1} \right). \quad (2)$$

The resistance in Eq.1 consists of three terms, where $\rho_0(V_G)$ is the residual resistance at low temperatures, $\rho_A(T)$ the contribution due to acoustic phonon scattering in graphene, and $\rho_B(V_G,T)$ the contribution attributed to RIP scattering, consisting of terms for the SiO$_2$ substrate and an added term for the SCO nanoparticles. Furthermore, $B_1$ and $\alpha$ are fit parameters, and $k_B$ is the Boltzmann's constant. The values of 59 meV and 155 meV in Eq. 2 correspond to the energies of the two strongest optical phonon modes in SiO$_2$.[38] Considering an additional phonon mode for the SCO nanoparticles and a coupling term with the charge carriers, the added term in the resistance due to the nanoparticle (NP) layer $\rho_{B\ NP}$ can be written as (where *ss* indicates a spin-state dependence):

$$\rho_{B\,NP}(V_G,T) = e^{(-2.q.d)}.B_2 V_G^\alpha \left( \frac{g_3(ss)}{e^{\frac{E(ss)\,meV}{k_B T}} - 1} \right). \tag{3}$$

As the SCO nanoparticles are stabilized by organic polymers, the prefactor $e^{(-2.q.d)}$ accounts for the spacing layer between the nanoparticles and the graphene sheet. The presence of this spacing layer of thickness $d$ exponentially decays the scattering amplitude of the interfacial phonon modes of momentum $q$.[39] The shell thickness $d$ has been set to 1.5 nm, corresponding to the expected polymer length and the thickness of the graphene sheet to 0.4 nm. For graphene, the phonon mode momentum $q$ can be expressed as:[40]

$$q = k_F = \sqrt{\pi.n(V_G,T)}, \tag{4}$$

where

$$n(V_G,T) = \frac{C.(V_G - V_{Dirac}(T))}{e}. \tag{5}$$

Here, C = 1.38 × 10$^{-8}$ F.cm$^{-2}$ is the gate capacitance per unit area, $V_{Dirac}$(T) is the Dirac point gate voltage and $e$ is the elementary charge.

In Eq.3, $E(ss)$ represents the spin-state dependence of the energy of the SCO nanoparticle phonon modes. For them, we use literature values of phonon mode energies of similar compounds measured by nuclear inelastic scattering found to be in the range of 37-62 meV and 25-37 meV in LS and HS state, respectively. We averaged these values in the model to 49.5 and 31 meV,[41] respectively (see Supplementary Fig. 17a).

The coupling of the SCO nanoparticle phonon modes with the graphene charge carriers is described through the term $g_3(SS)$, which amongst others, is dependent on the dielectric permittivity of the SCO layer.[37] The spin-state dependence of the dielectric permittivity as a function of frequency of the electric field has been reported by Lefter *et al.* while investigating powder samples of the same compounds. They found that the SCO compound polarizability drops 5 times when switching from the LS to the HS state.[36] We have taken into account such a memory effect in the model using their values (see Supplementary Fig. 17a). We find that the dielectric permittivity drop in the HS state yields a competing trend compared to the lower energy of the phonon modes, favouring respectively the observed lower resistance and a resistance increase. In other words, our model appears to indicate that the coupling between the SCO and the graphene resistance is mediated mainly by the changing electric polarizability of the compound, which is consistent with the recent observations of Zou *et al.*,[37] who reported an increase of the resistance of single-layer graphene after covering the graphene by a dielectric with a higher relative permittivity.

Figure 5 presents a comparison of the experimental results and model calculations, where the solid and dashed lines indicate the heating and cooling mode, respectively. The model captures the main observed experimental behavior, including the clockwise direction of the hysteresis loops. Moreover, the gate voltage dependence of the graphene sheet resistivity at room temperature is fairly reproduced, as well as the relative change of the resistance in both spin-states.

## Conclusions

In summary, we have demonstrated a new concept to detect, in a non-invasive way, the spin-state switching of a thin film containing nanoparticles made of the SCO compound [Fe(Htrz$_2$)(trz)](BF$_4$), and for a chemical alloy [Fe(Htrz)$_{2.95}$(NH$_2$trz)$_{0.05}$](ClO$_4$)$_2$ nanoparticle system. The method consists of measuring the electric properties of graphene in a four probe field-effect configuration, while covered with a monolayer of switchable nanoparticles. The coupling between the spin state dependent physical properties and the scattering in the graphene layer is ascribed to a changing contribution of remote interfacial phonon scattering. Using model calculations we showed that the charge carrier-nanoparticle coupling changes due to the spin-state dependence of the dielectric constant. As the contact printing method yields a homogeneous coverage of a thin film with single nanoparticle thickness, this graphene field-effect sensor device is already able to probe small volumes of *ca.* 2000 $\mu m^3$ in the experimental setup presented here, while exhibiting a clear spin-change dependence in the graphene resistance. In the near future this method should pave the way for the investigation of even smaller amounts of spin crossover material using optimized graphene nanosensor platforms. In addition, the simplicity of the sensor and its large operating temperature range promise flexible implementation of various other compounds that change their properties via external stimuli such as temperature or light.

# Experimental section

## Synthesis of spin-crossover nanoparticles.

Synthesis of systems 1 and 2 were described in detail in ref. 29.

## Sample preparation.

Devices were prepared using commercially available CVD-grown single-layer graphene on Si(backgate)/SiO$_2$(285 nm) substrates (obtained from Graphene Supermarket). Before characterization, the edges of the graphene sheet were mechanically removed to make sure there was no electrical contact between the graphene sheet and the Si backgate. Prior to deposition, the devices were annealed in N$_2$ at 380 K for 1 h and ramped several times from room temperature to this temperature until the charge neutrality point remained stable. Subsequently, thin layers (mostly monolayers) of SCO nanoparticles were deposited on top of CVD graphene surfaces using a micro-contact printing method, and deposition protocols described in our previous works.[13,27]

## Electrical transport measurements.

The conductivity was determined from the four-probe sample resistance using $\sigma = (L/W)(1/R)$, where $L \sim W$ in our configuration. It should be noted that there is some uncertainty in the geometrical factor L/W ($L \sim W \sim 350 \pm 25\ \mu m$). This translates into an uncertainty of *ca.* 14 % in the vertical axes of resistance and mobility data. As this is a systematic error, this does not change our conclusions. The data consists of measurements of the resistance as a function of gate voltage (V$_G$), at different temperatures. The R(V$_G$, T) curves are fitted using the diffusive transport field-effect model to extract device characteristics taking into account as a free parameter an asymmetry of the R(V$_G$, T) curves.[31,32] For the fits, V$_G$ is displayed from 0 to 220 V to evidence the ambipolar behaviour, as the Dirac point was experimentally found to be close to the upper limit of the gate voltage (111 V). The charge carrier mobilities were calculated by means of a least-squares linear fit to the steepest regime in the $\sigma$(V$_G$) for every temperature step, as described in reference.[42] Following their procedure, the mobility is calculated as $\mu = (1/C)\ d\sigma/dV_G$ and averaged over a $\pm$ 2 V interval around the steepest point of the derivative of the conductivity. The back-gate capacitance C corresponds to the SiO$_2$ layer capacitance, that is $C = \varepsilon \cdot \varepsilon_0 / t_{ox}$, where $\varepsilon$ and $t_{ox}$ are the relative permittivity and the thickness of the SiO$_2$ insulating layer equal respectively to $\sim$3.9 and 285 nm. Note, that the quantum capacitance of graphene was neglected since it is much larger than the geometric capacitance.

Temperature cycles were carried out using a heater element embedded in the probe station with a Lakeshore temperature controller and a local calibrated thermistor (TE-tech, MP-3011). Heating was performed at a rate of 5 K.min$^{-1}$ while the cooling rate is not constant but equal or below 5 K.min$^{-1}$ depending on natural heat dissipation rate.

Tungsten continuously variable temperature (CVT) flexible tips of 25 $\mu$m in diameter were used for their capability of compensating for thermal expansion of the probe arms. Temperature range from room temperature up to 385 K ensured a reliable device contact with CVT tips on graphene throughout several temperature cycles, which means that the tips remained in contact with the graphene for all temperatures.

## Characterization.

SCO nanoparticles were visualized after chemical synthesis by transmission electron microscopy (TEM) and after electrical measurements by optical microscopy and AFM. TEM visualization of the SCO nanoparticles was performed using a FEI Titan microscope operating at 300 keV. Scanning TEM (STEM) imaging was used for two reasons: to get a better contrast compared to standard TEM and to reduce the damage by the electron beam. Using high-angle annular dark field (HAADF) detector a high contrast was achieved due to enhancement of mass-thickness contrast, which leads to sufficient difference between high intensity nanoparticles separated by almost no-intensity gaps.[27] Size distribution analysis was performed using the ImageJ software.[43] AFM scans were acquired using a Bruker Dimension FastScan microscope in peakforce tapping mode in ambient conditions. Magnetic susceptibility measurements were performed on single-phased polycrystalline samples of 1 and 2 with a Quantum Design MPMS-XL-5 SQUID susceptometer. The susceptibility data were corrected from the diamagnetic contributions as deduced by using Pascal'âĂŹs constant tables. The data were collected in the range 300-400 K upon recording several heating-cooling modes at a constant rate with an applied field of 0.1 T.

## Acknowledgements

We acknowledge fundings from the EU (Advanced ERC grants SPINMOL and Mols@Mols), the EU project Graphene Flagship (Contract No. 604391) the Spanish MINECO (grant MAT2014-56143-R and Unit of Excellence Maria de Maeztu MDM-2015-0538) and from the Generalidad Valenciana (PROMETEO and ISIC Programs of excellence) for financial support of this work. M.G.-M. thanks the EU for a Marie Sklodowska-Curie postdoctoral fellowship (H2020-MSCA-IF-658224). T.K. thanks ERC project 267922 for support.

## Supporting Information Available

Additional experimental results are available. (1) STEM-HAADF images of SCO NPs, (2) SCO NP size distribution. (3) AFM characterization after SCO NPs deposition on graphene. (4) Magnetic properties of powder samples of SCO NPs. (5) Table summarizing extracted physical parameters of fitted data. Complementary data of gate/temperature dependence of graphene electrical properties before (6-7) and after (8-12) SCO NP deposition. (13) Simulated memory effect in the relevant physical properties.

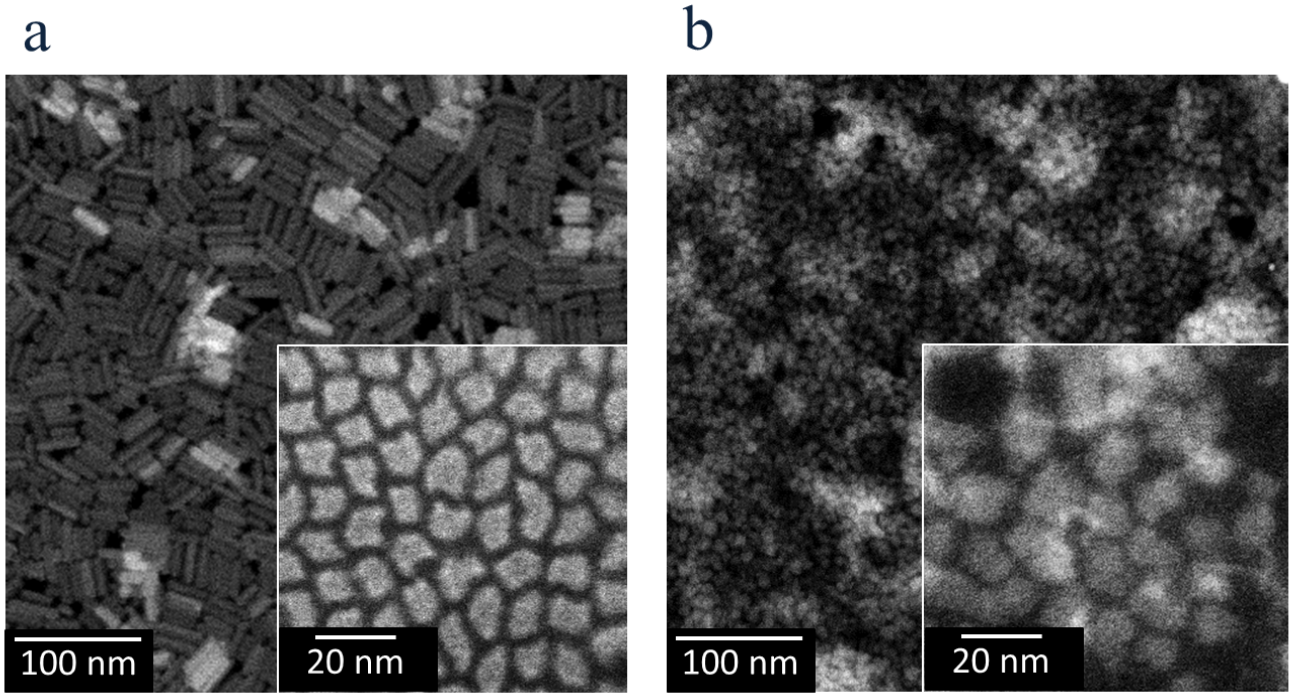

Figure 6: STEM-HAADF images of free-standing 2D sheets of SCO nanoparticles prepared by stamping technique. Insets show lateral self-assembly of nanoparticles obtained by a drop-casting method. **(a)** nanoparticles of system (1) ($\Phi$ = 9 nm and L = 25 nm) and **(b)** nanoparticles of system (2) ($\Phi$ = 10 nm).

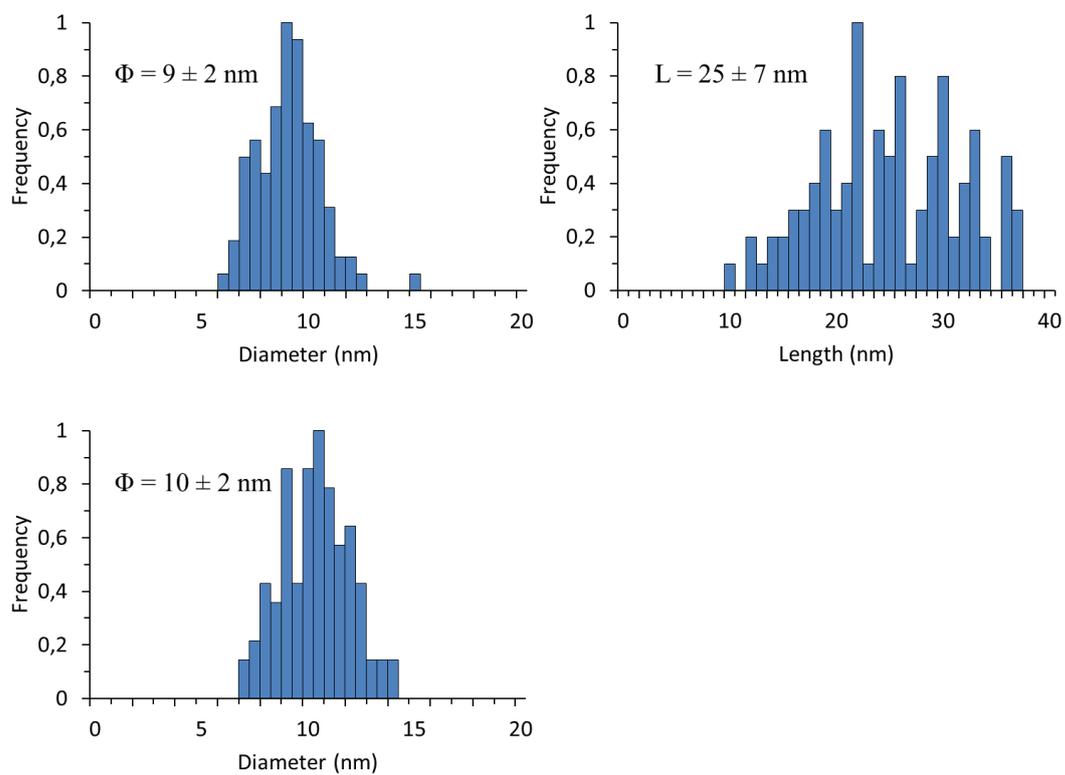

Figure 7: Obtained mean dimensions and standard deviation for nanoparticles of system (1) [Fe(Htrz)$_2$(trz)$_1$](BF$_4$) [diameter (top left) and length (top right)] and system (2) [Fe(Htrz)$_{2.95}$(NH$_2$trz)$_{0.05}$](ClO$_4$)$_2$ [diameter (bottom)] from TEM images (100 counts): system (1) Diameter Φ = 9 ± 2 nm and Length = 25 ± 7 nm; system (2) Diameter Φ = 10 ± 2 nm.

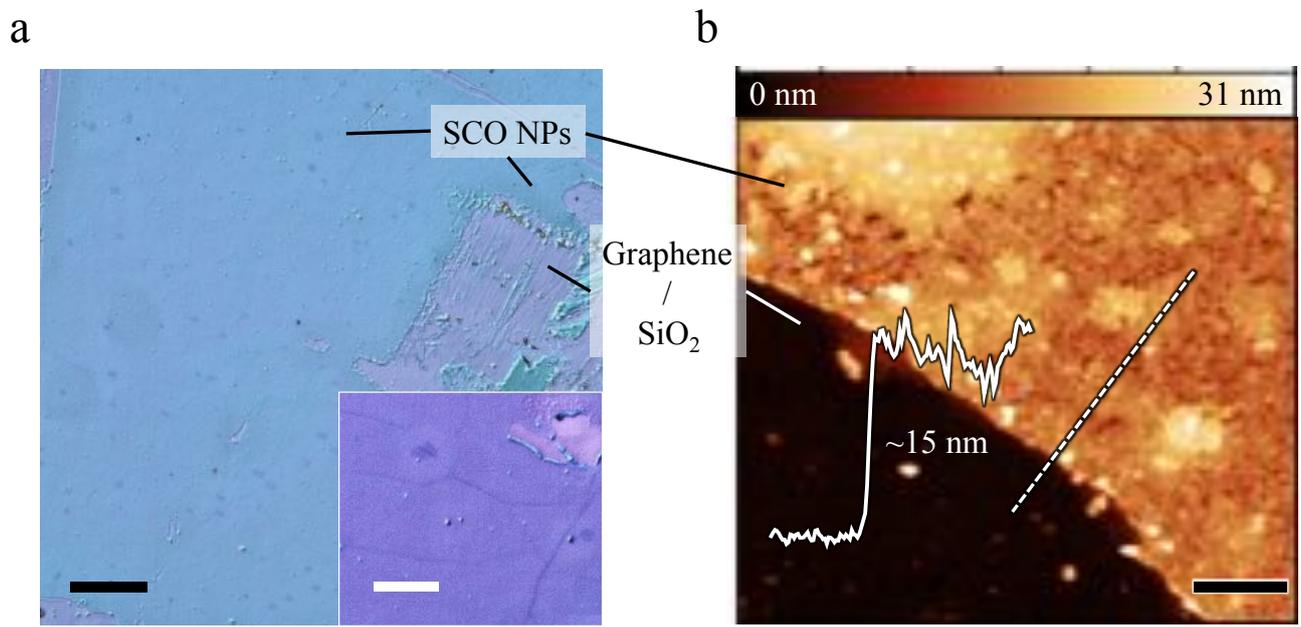

Figure 8: (**a**) Optical image of the sample, where a thin film of the SCO compound [Fe(Htrz)$_{2.95}$(NH$_2$trz)$_{0.05}$](ClO$_4$)$_2$ is printed on single layer CVD graphene. Scale bars correspond to 50 and 10 $\mu$m respectively for the whole image and in the inset. (**b**) AFM image of the SCO thin film, showing a step of *ca.* 14.8 $\pm$ 1.7 nm fairly close to one nanoparticle averaged diameter. The scale bar correspond to 1 $\mu$m.

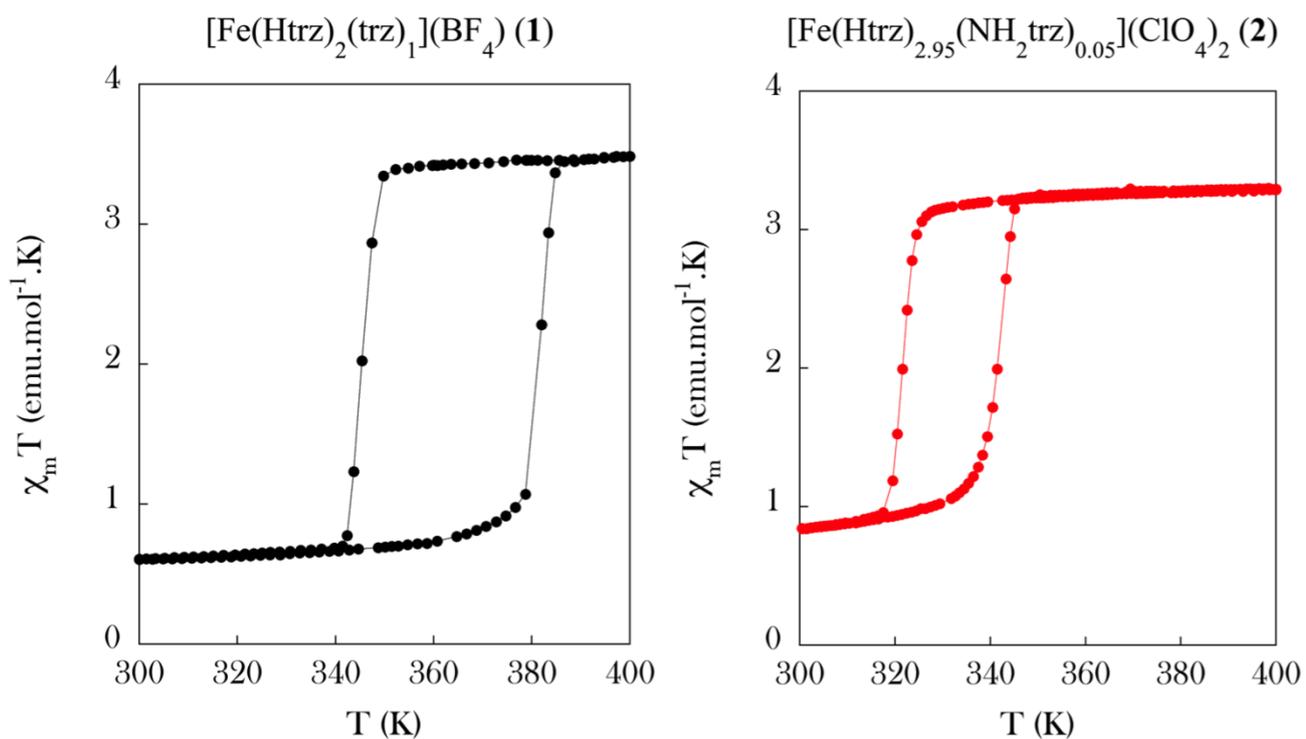

Figure 9: Temperature dependence of the molar magnetic susceptibility temperature product ($\chi_M.T$). **a,** and **b,** present respectively data for system (1) (25*9 nm) and system (2) (10 nm), after several heating–cooling modes recorded in the temperature range 300-400 K under an applied magnetic field of 0.1 T (heating/cooling rate of 1 K.min$^1$).

Table 1: Summary of the extracted physical parameters of fitted data presented in Fig.10. The mobility of cycle 1 should be seen as an indication only, as the Dirac point was out of the experimental scope.

| Cycle | T (K) | Atm | $\mu_h$ (cm$^2$/(V.s)) | $\mu_e$ (cm$^2$/(V.s)) | $\mu_e/\mu_h$ | $R_{Dirac}$ (k$\Omega$.sq$^{-1}$) | $V_{Dirac}$ (V) |
|---|---|---|---|---|---|---|---|
| 1 | 305 | air | 273 | - | - | $> 0.7$ | $> 85$ |
| 2 | 305 | N$_2$ | 1127 | 903 | 0.80 | 5.6 | 59.0 |
| 3 | 305 | N$_2$ | 1402 | 1231 | 0.88 | 3.3 | 27.6 |
| 4 | 380 | N$_2$ | 938 | 635 | 0.68 | 4.3 | 33.4 |
| 5 | 305 | N$_2$ | 1378 | 1104 | 0.80 | 3.3 | 28 |
| 6 | 380 | N$_2$ | 1114 | 710 | 0.64 | 4.4 | 26 |

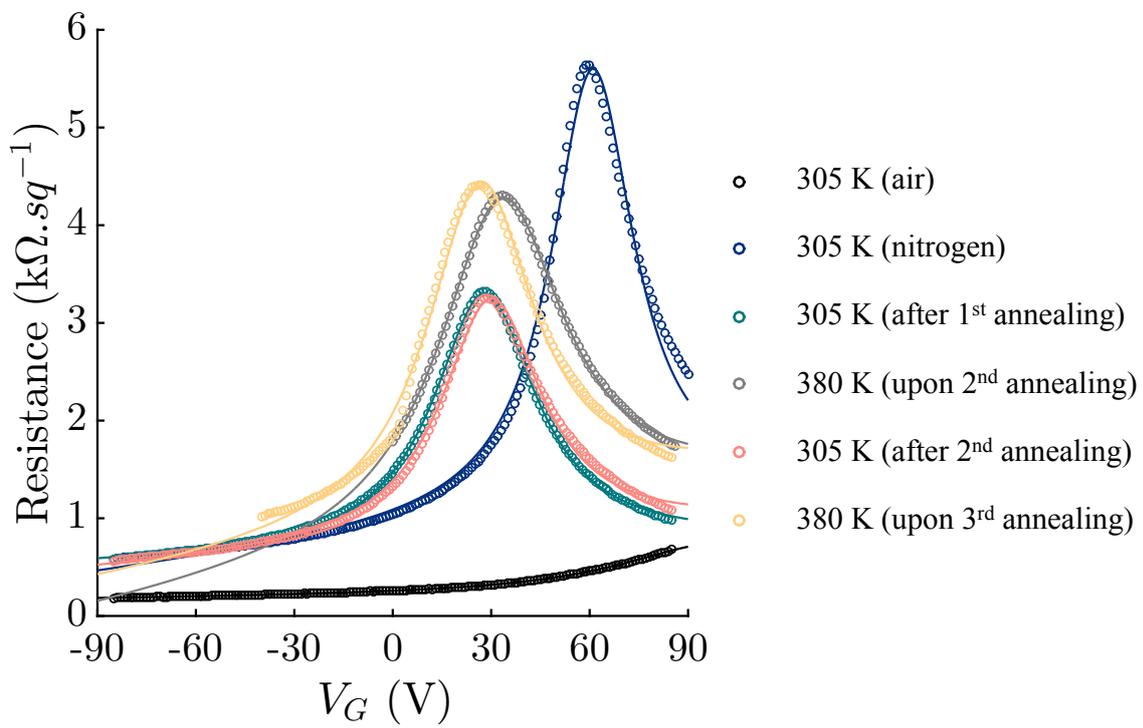

Figure 10: Gate-dependent electrical properties of the graphene before spin-crossover nanoparticle deposition. Dots and lines correspond respectively to data and model fits for different annealing steps. Flushing of nitrogen shifts the Dirac point within the experimental range, and further heating (maintaining $N_2$ flow) up to 380 K brings it closer to zero gate voltage. This shift stabilizes after the $2^{nd}$ annealing cycle.

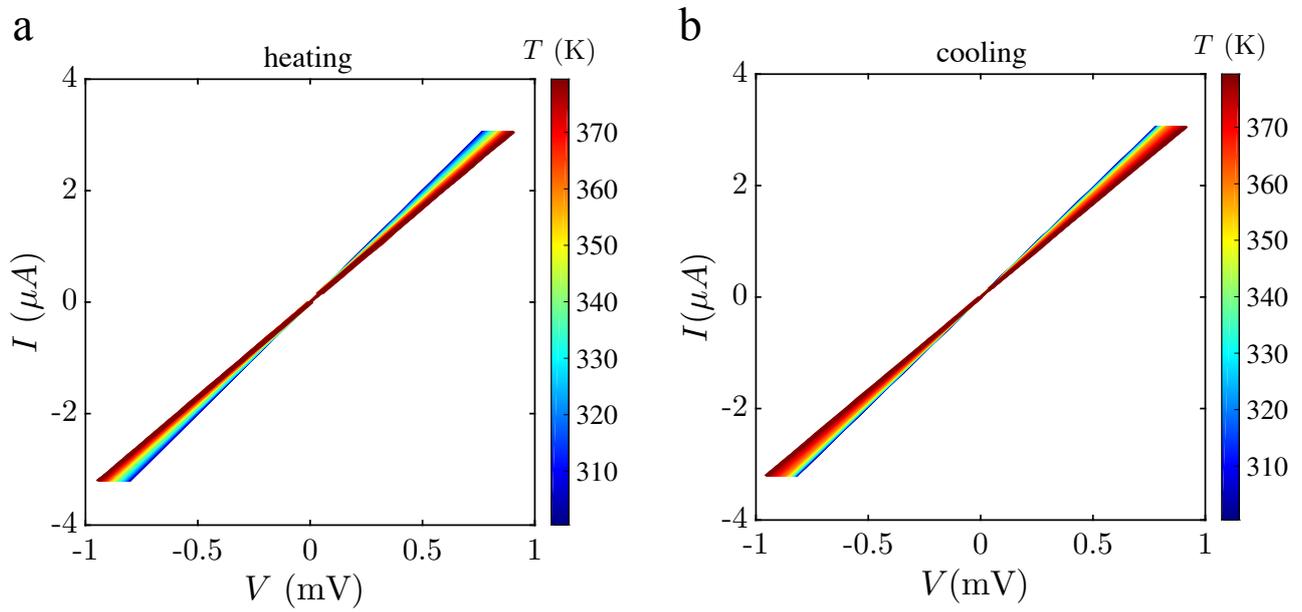

Figure 11: Temperature dependence of the Current-Voltage (IV) characteristics in heating (**a**) and (**b**) cooling modes at zero gate voltage recorded after the graphene was cleaned and pre-conditioned. The linear behavior of the IV characteristics upon heating mode indicates an ohmic contact for the electrodes on graphene.

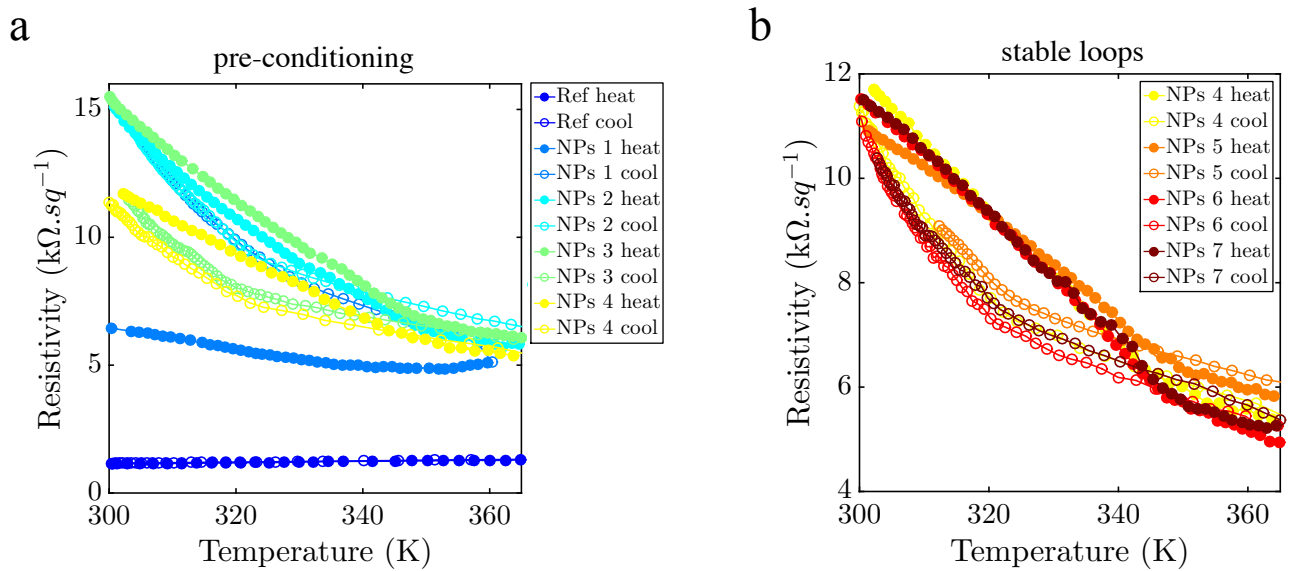
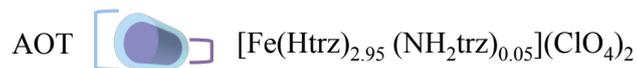

Figure 12: **a and b,** display resistance data at zero backgate voltage as a function of heating and cooling modes measured under $N_2$ atmosphere (full and empty dots, respectively). (**a**) The dark blue dots show the behavior of the resistance upon a thermal cycle for a device measured before deposition of the nanoparticles on top of the graphene sheet (*i.e* the reference of the study). The reference shows the expected metallic behavior (increase of the resistance with the temperature) far away from the Dirac point ( 30 V). Multiple thermal cycles (1-4) measured after deposition of the nanoparticles are shown. A substantial increase of the resistance occurs for the first 3 cycles then stabilizes at the cycle 4, which exhibits a hysteresis loop in the thermal resistance. (**b**) A change in the slope of the resistance is reproducibly observed above(below) 345( 320 K) in the heating(cooling) mode for the next three temperature cycles.

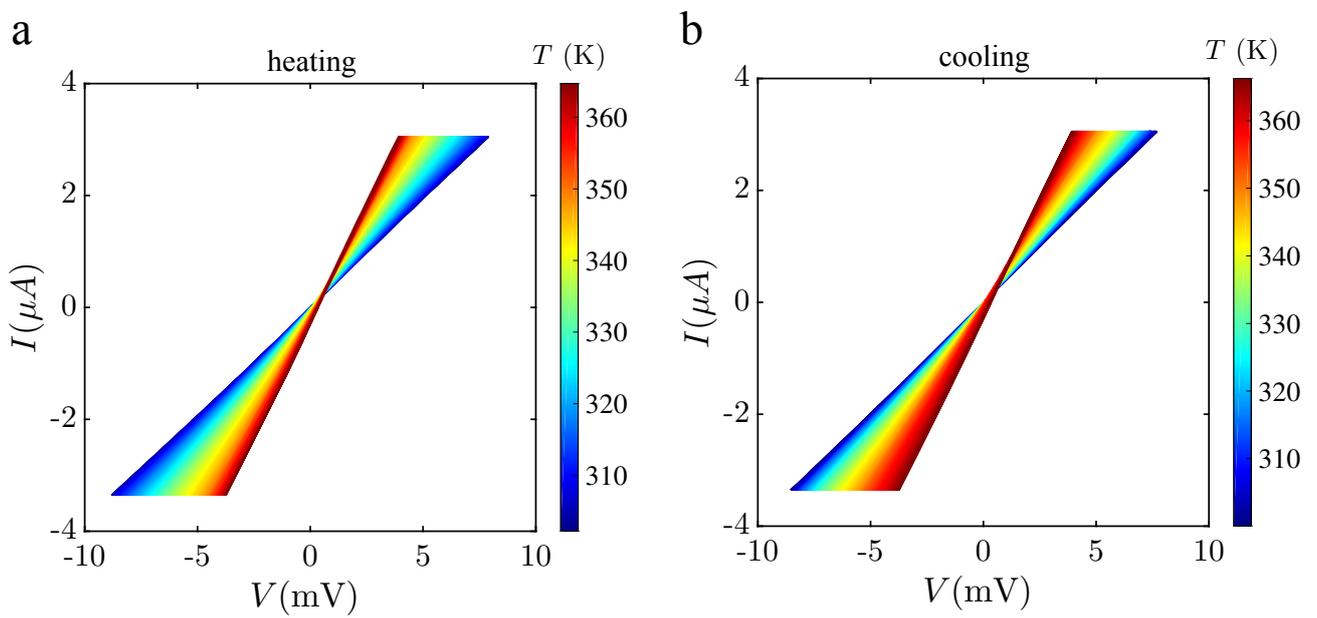

Figure 13: Temperature dependence of the I-V characteristics in heating (**a**) and cooling (**b**) modes at zero gate voltage recorded, after the graphene was cleaned, pre-conditioning and after nanoparticle deposition of the NP system (2) (corresponding to the first stable cycle number four). The linear behavior of the IVs upon heating mode indicates an ohmic contact for the electrodes on graphene through the nanoparticle thin-film.

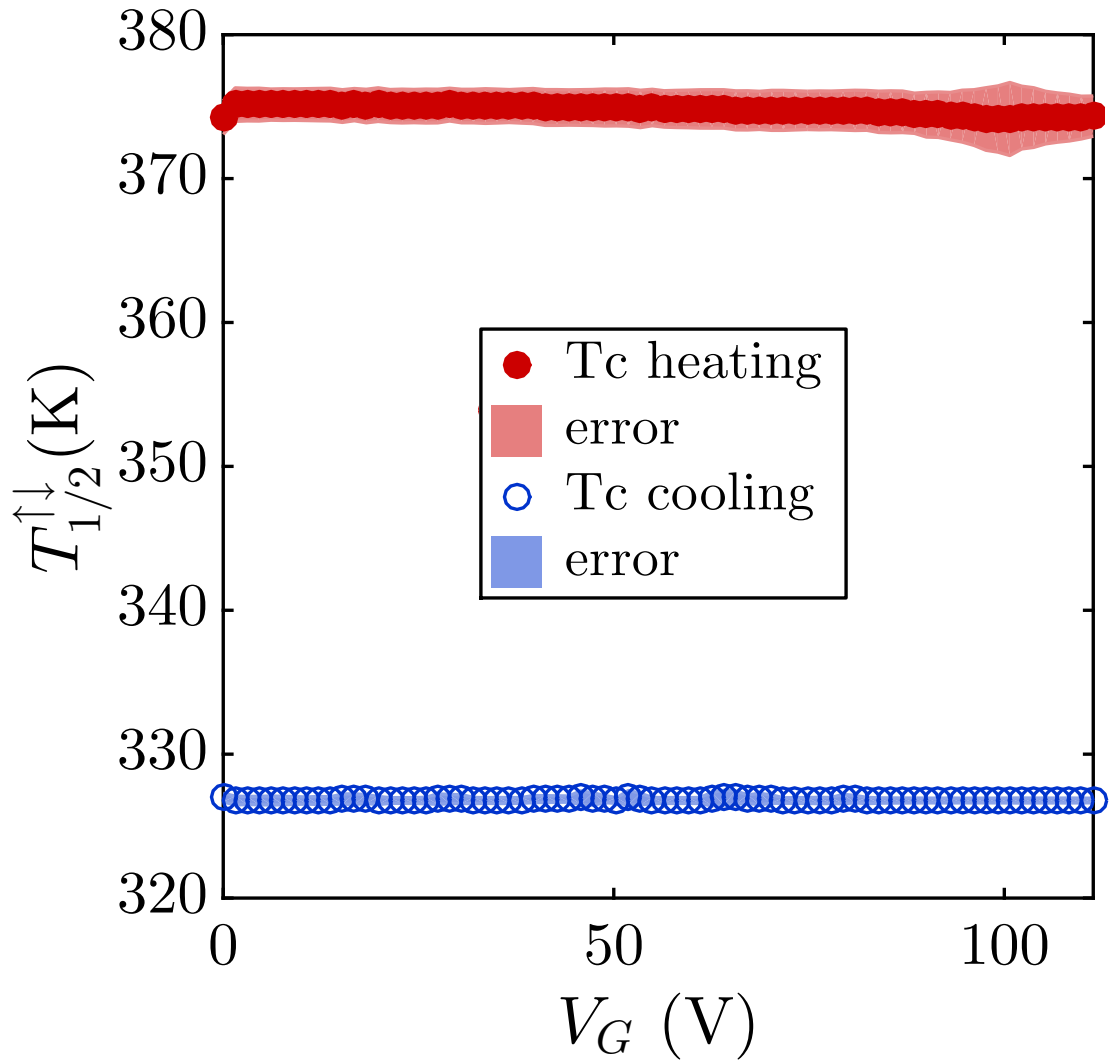

Figure 14: Critical temperatures of the spin-transition of the SCO nanoparticle system (1). Extracted values from the maximum of the derivative of the resistance versus temperature at different backgate voltage (see Fig. 3.b) fitted by a Gaussian function. The vertical shaded error bars are the FWHM of the Gaussian fit. This graph evidences the non-backgate dependence of the critical temperatures of the spin-transition.

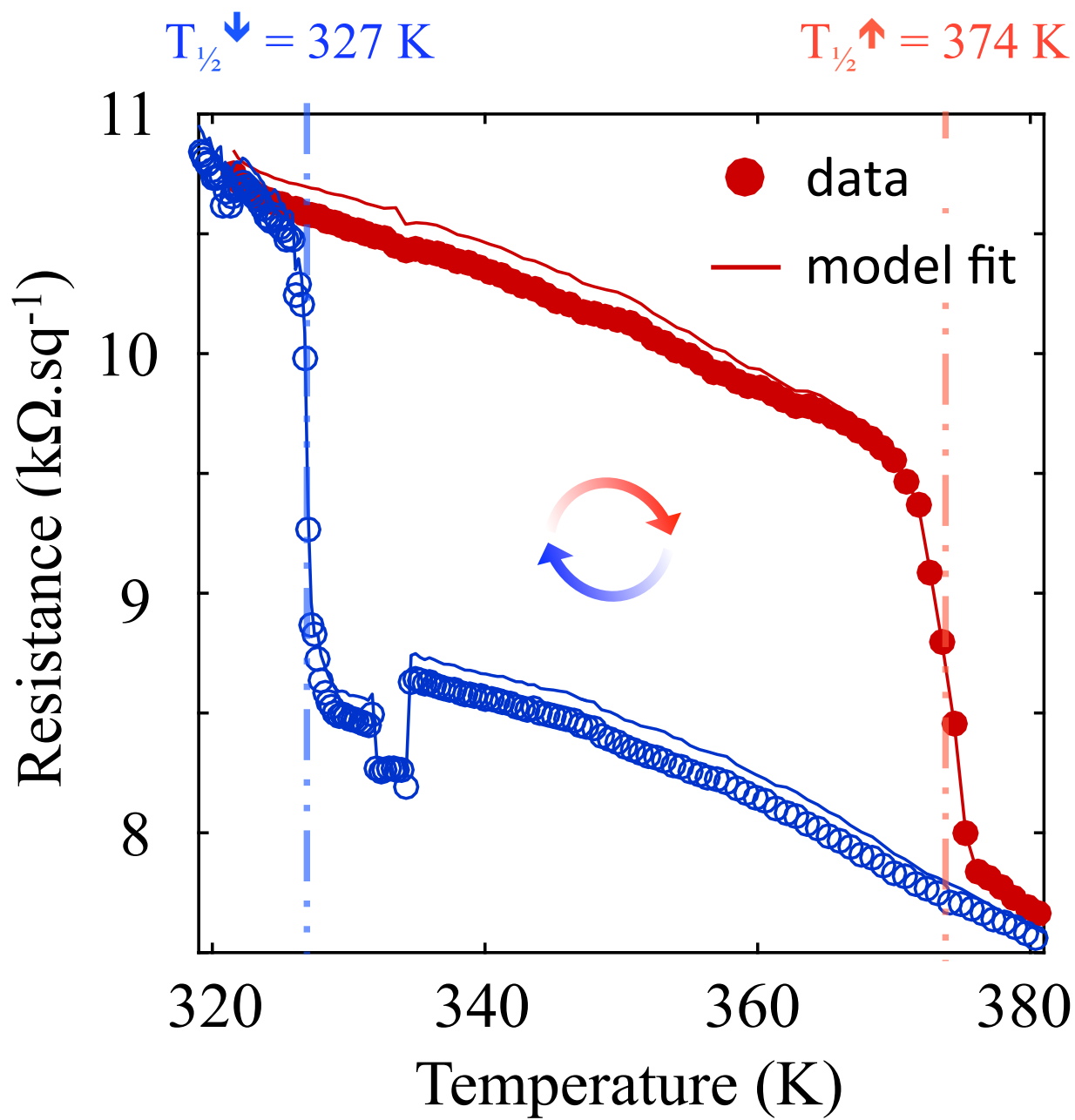

Figure 15: Resistance per square at the Dirac point as a function of temperature for heating and cooling modes after nanoparticle deposition of system (1). Values from experimental data and model fit are denoted by dots and solid lines, respectively. A wide clockwise hysteresis loop is observed between 327-374 K.

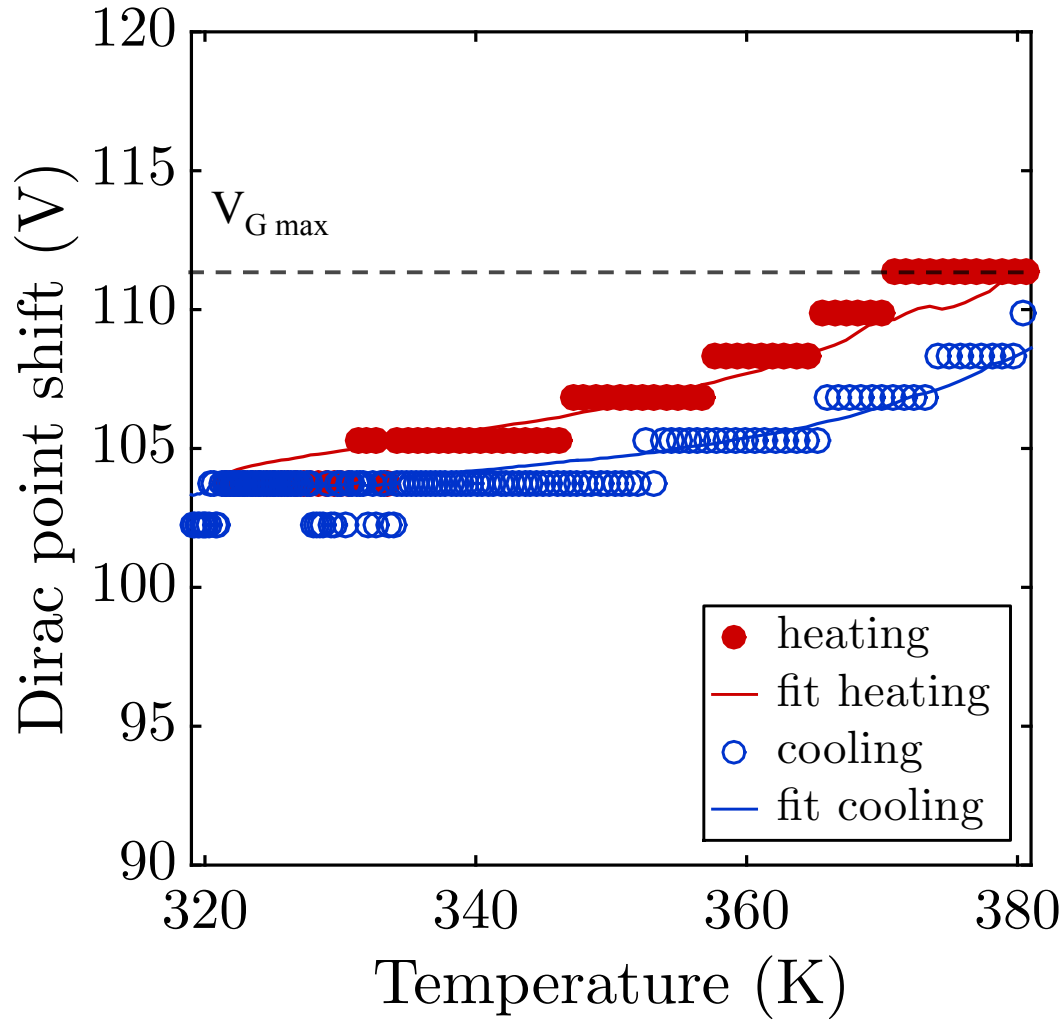

Figure 16: Temperature dependence of the Dirac point voltage for heating and cooling modes after nanoparticle deposition of system (1). A continuous increase is seen, up to the upper limit of the experimental gate voltage ($V_{G\,max}$, denoted by the dashed line). Importantly, the heating/cooling modes almost match and do not evidence any abrupt transition around the expected hysteresis loop temperatures, Illustrating that there is no spin-state dependent charge transfer between the nanoparticles and the graphene. The incremental increase in the data points (symbols) and the saturation of the Dirac voltages observed close to 380 K, is respectively due to the backgate voltage resolution of ∼1.5 V and the fact that the Dirac voltage shifts above $V_{G\,max}$, the highest sweeping experimental value of backgate voltage applied.

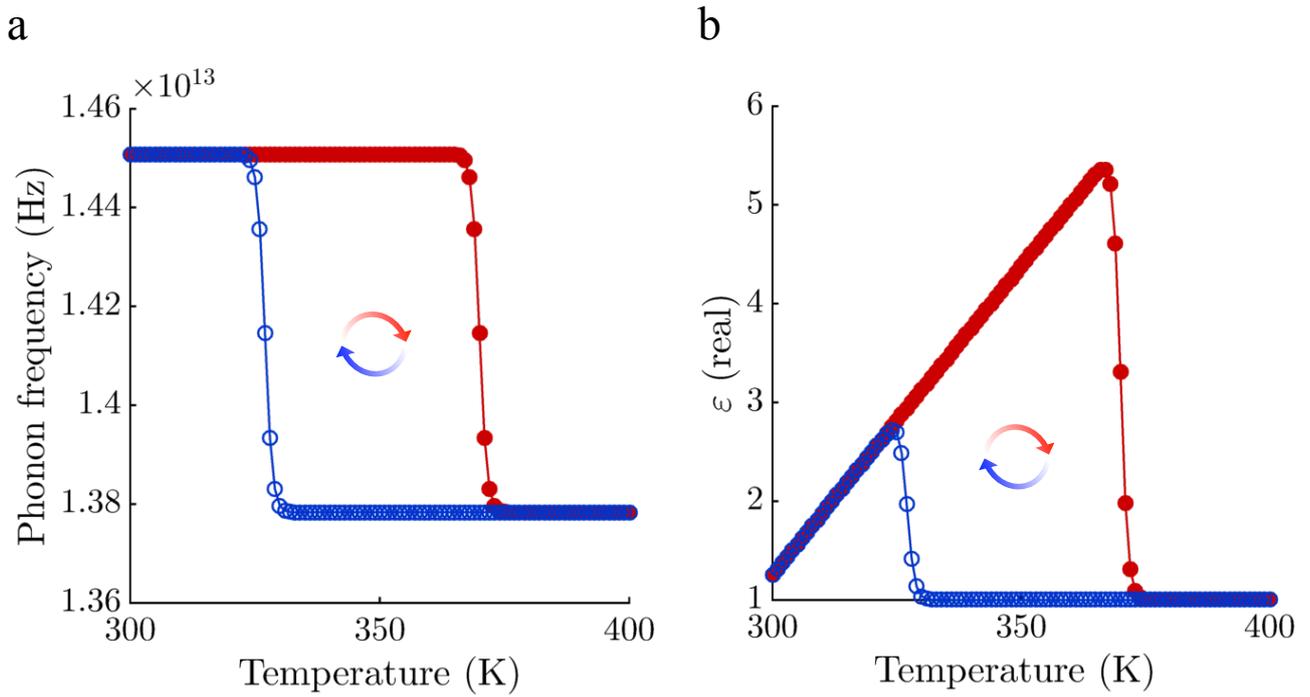

Figure 17: Simulated thermal hysteresis loop of the nanoparticle system (1) relevant physical properties. (**a**) phonon frequency and (**b**) real part of the dielectric permittivity. For the former case, we used literature values of phonon mode energies of similar compounds measured by nuclear inelastic scattering found to be in the range of 37-62 meV and 25-37 meV in LS and HS state, respectively. We averaged these values in the model to 49.5 and 31 meV,[41] respectively. For the latter case, we used the real part of the relative permittivity recorded at 1 KHz by Lefter *et al.*[36] for the same SCO compound than the one used here for comparison between experimental results and model calculations.

# Graphical TOC Entry

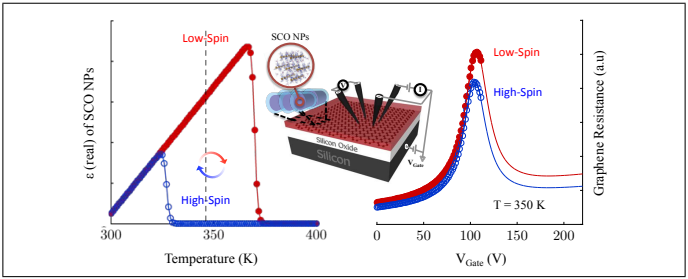